\documentclass{ws-rv9x6}
\makeindex

\newtheorem{principle}{Principle}[chapter]

\begin{document}

\chapter{Thoughts on the Cosmological Principle}

\author[Dominik J.~Schwarz]{Dominik J.~Schwarz}

\address{Fakult\"at f\"ur Physik, 
Universit\"at Bielefeld, Postfach 100131, 33501Bielefeld,  Germany}

\begin{abstract}
\end{abstract}

\body

\section{Introduction} 

Wolfgang Kummer was a great teacher and mentor to me. Although Wolfgang never published any research on cosmology himself, he had a great interest in that field and supported me in my decision to get engaged with cosmological issues. Thus, I decided to describe current ideas and some of my own thoughts on one of the corner stones of modern cosmology --- the cosmological principle. This principle says that the Universe is spatially homogeneous and isotropic. It 
predicts, among other phenomena, the cosmic redshift of light, the Hubble law and the black body shape of the cosmic background radiation spectrum. Nevertheless, the existence of
structure in the Universe violates the (exact) cosmological principle. A more precise  
formulation of the cosmological principle must allow for the formation of structure and must therefore 
incorporate probability distributions. 
Below, I discuss how to formulate a new version of the cosmological principle, how to test it, and how to possibly justify it by fundamental physics. But let me, before doing so, describe in short some of my memories of Wolfgang. 

\section{Tribute to Wolfgang Kummer} 

My first contact with theoretical physics was with Wolfgang. He taught a course on ``Methods in Theoretical Physics'', which was compulsory for all physics students at the Vienna University of Technology (TU Wien) during their second year. The exercises accompanying that course were demanding and I learned how to handle complicated calculations. Later on I enjoyed his excellent lectures on ``Particle Physics'', which
triggered my decision to devote my studies to fundamental issues in physics and to seek for a 
possibility to become part of his research group at the Institute for Theoretical Physics (ITP). 

Luckily, during the third year of my study, the position of a library assistant was vacant in the ITP and Wolfgang was looking for a student interested in that position. So I became a member of Wolfgang's group well before I started my diploma project.  The duties of a library assistant occupied not more than one or two hours a day and  I was able to concentrate on my studies and research. Consequently, 
Wolfgang became my diploma advisor. The diploma research project was on two-dimensional 
gravity in the context of Riemann-Cartan geometry. Wolfgang had already done some preliminary, unpublished work, which provided a good starting point. At the same time, there 
appeared a very interesting work by Katanaev and Volovich, which opened up interesting 
perspectives. Wolfgang's style to approach a new problem and his attitude to meet his students at 
an equal level impressed me very much and is still influencing me in the way I try to deal with my students. We met every week to develop new ideas and to check all calculations step by step and soon 
managed to quantise the system and to find all its classical solutions. (For the scientific aspects of our work, see the contribution of L.~Bergamin and R.~Meyer to this volume.) 
This work resulted in my diploma thesis, three publications and several proceeding articles.

After my diploma thesis, I decided that I would like to devote my research towards a topic closer to ``experiment'' and chose cosmology. My first contact with modern cosmology was probably in the weekly theory seminar of Wolfgang's group, when we worked through the book of Kolb and Turner.
Wolfgang took care that all graduate students of the ITP would have the chance to participate at international conferences and workshops. During one of those, which I was lucky to attend during my PhD studies, the COBE discovery of cosmic temperature anisotropies was presented for the first time 
in Europe.  A big tradition at the ITP was and is the study of gauge theories (starting off from Wolfgangs important contributions on axial gauges, see the contributions of P.~Landshoff, D.~Blaschke et al., and 
P.~Landshoff and A.~Rebhan in this volume), 
and in the 1980s there was some confusion in the cosmology community on the issue of whether and how to use ``gauges'' in cosmological perturbation theory. Anton Rebhan picked up that topic and combined the cosmological perturbation theory with techniques and methods from  finite temperature field theory, which attracted me to become Toni's PhD student. To my surprise, Wolfgang supported my decision to change the field, while he continued the study of two-dimensional physics. He became my mentor and continued to support my career. Besides our scientific connections, Wolfgang and Lore Kummer have been good friends to me and my family. I am very thankful for his support and will keep Wolfgang in good memory.

\section{Modern Cosmology}

Although questions of cosmology have been an issue for thousands of years,  
only the 20th (Christian) century saw cosmology turning into a physical science. 
Einstein's general relativity allows us to talk about the space-time of the Universe and to 
formulate dynamical laws for its geometry. Light is our most important source of information 
to learn about the evolution and state of the Universe. The advent of quantum mechanics, atomic, nuclear and particle physics enabled us to understand the mechanisms of light emission and
absorption. At the same time, astronomical observations became sensitive, numerous and 
precise enough to study the global properties of the Universe. 

Inspired by the ideas of Mach, Einstein decided to select very special conditions for a model of the Universe. His first attempt was to find solutions to his equations that allow for a static and 
spatially homogeneous and isotropic space-time. In order to achieve that, he had to introduce an additional term to his equations --- the cosmological constant. With Hubble's discovery of cosmic expansion Einstein realised that the Universe was not static and he dropped the cosmological constant. 
This opened the way for the success of the Friedmann-Lama\^itre models, which are characterised by  spatial homogeneity and isotropy. Milne coined the name ``Cosmological Principle'' for the statement
that these symmetries are realised (at least approximately) 
in the Universe (see Peeble's book \cite{Peebles} for a more detailed description of the history of these ideas). 
Today, we have reached a high level of precision and as the cosmological principle is at its best an approximate statement about Nature, 
it is timely to think about possible refinements, especially in the light of the recently discovered cosmic 
acceleration of the Hubble expansion. In that context it has been proposed that the apparent cosmic 
acceleration might be an inappropriate interpretation of the data, due to our ignorance with respect to the effect of averaging over cosmic distances \cite{backreaction}. 

\section{Observational Facts} 

Typical cosmic photons belong to the cosmic microwave background (CMB) radiation. 
The observation of these microwave photons provides the basis of modern cosmology. 
The CMB radiation is well described by a black body at temperature $T_0 = 2.7$ K, is 
almost perfectly isotropic over the full sky and almost unpolarised. A small dipole anisotropy  
$\Delta T/T \simeq 10^{-3}$ is interpreted to be due to the motion of the Solar System barycentre 
with respect to this cosmic heat bath. At smaller angular scales cosmic temperature anisotropies 
are tiny, $\Delta T/T \simeq 10^{-5}$. 

A high degree of isotropy is actually observed at all explored frequencies of the electromagnetic radiation (if one disregards nearby objects). Not only does the CMB have this property, but even 
the angular distributions of astrophysical objects on the sky at the extreme ends of the 
electromagnetic spectrum, radio galaxies and gamma-ray bursts, are isotropic. 

This suggests that the distribution of light in the Universe is statistically isotropic. This would imply, 
that the probabilities to see a supernova, to find a radio galaxy or to measure a certain amount 
of CMB polarisation are distributed uniformly on the sky. 
However, this statement is obviously violated by several local phenomena, like day and night,  
or the Milky Way. A potentially true statement is:

\begin{proposition}[Statistical Isotropy]
Apart from anisotropies of local origin,
the distribution of light in the Universe is statistically isotropic 
with respect to the barycentre of the Solar system.\footnote{We could also refer to the barycentre of the Milky Way or of the Local Group, but those are less well known and it would not solve the problem that there might still be unresolved local effects.} 
\end{proposition}

Local origins of anisotropy are, e.g.~the Zone of Avoidance caused by the Milky Way, 
or the motion of the Solar system barycentre with respect to the CMB.

Causality is a fundamental principle of modern physics. However, it does not play any role when  discussing the issue of statistical isotropy. This is no longer the case when we discus the question of 
spatial homogeneity. Our observations allow us to estimate distances of objects that are located on our backward light cone. Thus looking at distant objects means that we are also looking back in time. 
This is a substantial complication, as it means that we cannot study the issue of spatial distributions without a model of cosmic evolution. 

The three dimensional distribution of matter in the Universe is observed by means of redshift surveys. 
Studying their distribution, we first of all find that galaxies come in groups, clusters and super-clusters. There exist big voids surrounded by filaments and sheets of structure. The largest object found in the Universe so far is the Sloan Great Wall, which extends over a few 100 Mpc \cite{Gott}.

In a static or stationary Universe we could expect homogeneity in redshift space, but as we know that there is evolution in the Universe (e.g., the ratio of ellipticals to spirals changes as a function of redshift, 
the ionisation of the intergalactic medium changes at a redshift of $z \simeq 6$, \dots), there cannot be 
homogeneity in redshift space. However, in an evolving Universe it does make sense to study the distribution of matter on spatial hypersurfaces,  their definition being observer-dependent. 

It seems useful to talk about the spatial hypersurface that is defined by a real astronomer. We might 
correct for some well understood effects, like the motion of Earth in the solar system. 
The astronomer can define her unique {\bf comoving spatial hypersurface}. Let me also note that the word comoving obviously has to refer to the motion of atomic matter here. In general relativity one usually defines a class of {\bf comoving observers}, which means that they are comoving with some form of matter. It seems feasible to define the class of {\bf atomic/baryonic comoving observers} 
(as it is possible to receive information from them, while I don't know a way to receive information from an observer made out of dark matter). In the following we will refer always to them.

A perfectly homogeneous distribution is characterised by a well defined mean density 
(one-point correlation) and the vanishing of the (reduced) higher n-point correlation functions. 
A volume independent mean density seems to exist on scales larger $100$ Mpc \cite{Hogg}, 
but this issue remains controversial \cite{Labini}. The vanishing of the two-point 
correlation at scales much larger $100$ Mpc is best seen by means of quasar redshift catalogues 
\cite{Q2dF}. Although it is not clear if statistical homogeneity does hold, we formulate

\begin{proposition}[Statistical Homogeneity]
The spatial distribution of visible matter in the Universe on scales larger than a 
homogeneity scale $r_h$ is statistically homogeneous.
\end{proposition}  

In the following we will always assume that proposition 1.1 holds true and investigate its implications. 

It is important to realise that the existence of a globally defined {\bf cosmic time} is closely related to the large scale homogeneity (proposition 1.2) of the Universe.  It implies that observers at different places in the Universe probe just different realisations of the same distribution of light and matter, which can be parametrized as a function of cosmic time. If statistical homogeneity does not hold, different observers might experience very different histories of the Universe. 

\section{Formulation(s) of the Cosmological Principle}

These observationally motivated propositions are usually combined with a statement that 
seems to be a logic continuation of Bruno's and Copernicus' insight that we do not live at 
the centre of the world. Let me formulate two different versions: 

\begin{principle}[Weak Copernican Principle] 
We are typical.
\end{principle}

\begin{principle}[Strong Copernican Principle]
We are not distinguished.
\end{principle}

The strong version is more radical. The weak version implies that typical observers, wherever they 
are and whoever they are, observe the same distributions. 

The strong version allows for different classes of observers, like there are different species of monkeys, none of them is distinguished. It is not a priori obvious that there couldn't be several species of observers, e.g.~those living in a spiral galaxy and those in an elliptical, or observers in a filament and observers in a void. These observers could observe statistically different distributions. 

What I call the weak Copernican principle is the commonly adopted textbook version. However,  
that we are made out of atomic matter, while the dominant mass/energy of the cosmic substratum 
seems to be non-atomic, questions the validity of the weak version. If we do not know these 
95\% of the Universe, how can we claim that we are typical? 

We can now proceed to the formulation of a cosmological principle. At that point one usually 
lifts the statistical isotropy and statistical homogeneity on sufficiently large scales to an 
exact isotropy and homogeneity of space-time itself. 
The justification is that the isotropy of the CMB is almost exact and that one can assume exact 
isotropy as a starting point for a theory of structure formation. Combining the exact isotropy 
around one point with the weak Copernican principle, one concludes that every observer sees an isotropic sky. Together with some technical assumptions on the smoothness of the space-time metric, exact homogeneity follows \cite{Thirring}.  

\begin{principle}[Cosmological Principle]
All physical quantities measured by a comoving observer are spatially homogeneous and isotropic.
\end{principle} 

This formulation leads us to the class of Friedmann-Lema\^itre models, which are successful in describing 
the cosmic expansion and the thermal history of the Universe,
especially primordial nucleosynthesis and the decoupling of light. However, these models do not explain how structure forms. In order to do so, we have to introduce cosmological perturbations, which violate the
cosmological principle. 

Note that the cosmological principle as usually stated is much stronger than what we can 
possibly establish by means of observations. At best, it is only the statistical distribution of matter and light that appears to be homogeneous and isotropic, not its actual realisation. I thus favour 
an alternative formulation of the cosmological principle. 

\begin{principle}[Statistical Cosmological Principle]
The distribution of light and matter in the Universe is 
statistically isotropic around any point, apart from anisotropies of local origin.
\end{principle}

The observed isotropic distribution of light (and matter) together with the weak Copernican principle 
implies the statistical isotropy around every point. It seems to me, that this implies statistical 
homogeneity, however, I am not aware of a rigorous proof of that statement. However, perhaps we should use the strong Copernican principle and then we cannot conclude that homogeneity holds 
true. In that case we could only state a

\begin{principle}[Minimal Cosmological Principle] There exists a class 
of observers that see a statistically isotropic Universe, apart from anisotropies of local origin.
\end{principle}
 
This is a very interesting possibility, as this is the minimal version that seems to be justified by experiment. I think that the study of it's implications would be very interesting and could lead us to 
conclusions that differ significantly form today's textbook cosmology. 

\section{Testing the Cosmological Principle}

As an approximation, the cosmological principle is very useful, but strictly speaking it is
wrong. With respect to isotropy, we know that the violation is small, however with respect to 
homogeneity the case remains unclear. 

This actually might be at the reason for the current crisis that we are facing in cosmology: we claim 
that we know with high precision that we only understand 5\% of the Universe \cite{WMAP5yr}. 
But we have no direct evidence for the existence of any dark matter or energy. 

This is one of the reasons why many groups started to investigate the idea of cosmic backreaction as an alternative to the existence of dark energy. Instead of looking at statistical distributions, we can average over regions of space-time and study the properties of these estimators. These regions might be one, two or three dimensional. Due to the non-linearity of gravity, it is obvious that these estimates of physical quantities and the evolution of physical observables do not necessarily commute. This could give rise to a misinterpretation of the data and thus the cosmic acceleration could be an illusion \cite{backreaction}. 

This finally leads us to the question how one could test the statistical cosmological principle. There are 
some indications that statistical isotropy is violated at the largest scales on the CMB 
\cite{CMBanomalies}, but it remains to be seen if that will eventually turn out as a Solar system contamination or a systematic effect. I mentioned already that the statistical homogeneity has not been firmly established so far.  

While all observations are consistent with the strong Copernican principle, its weak version is contradicted by our claim that the Universe is dominated by non-atomic stuff. 
This might be an irrelevant detail, thus several tests of the weak Copernican principle have been proposed \cite{CopernicanPTest}. 

\section{Cosmological Inflation and Quantum Gravity}

Can we justify the statistical cosmological principle? The scenario of cosmological inflation is certainly an important step towards a possible justification. In the context of eternal inflation \cite{einfl}, 
the classical version of the cosmological principle fails miserably at super-large scales, as the Universe is extremely inhomogeneous at these scales. In any case, it fails at scales larger than the particle horizon, which are enormously bigger than what we can observe and will ever observe. But, we can hope to justify the statistical cosmological principle for regions smaller than the particle horizon.

To sum up, the historically important formulation of the cosmological principle has no justification in modern cosmology, as the quantum fluctuations during inflation spoil it. However, turning it into a statement on the statistical distribution of light and matter seems to be a logic consequence of the very same quantum fluctuations. Unless a consistent formulation of quantum gravity is available, it seems 
that a cosmological principle of some form is still required.

The promises of quantum gravity to eventually predict the statistical cosmological principle also provide a link to Wolfgang's dedication to fundamental science--- the understanding of the quantum effects of space-time. 

\section*{Acknowledgement} 

I would like to thank the editors Daniel Grumiller, Toni Rebhan and Dimitri Vassilevich for 
their effort to put together this memorial volume.

\end{document}